\documentclass[preprint,showkeys]{revtex4-2}

\usepackage{graphicx}
\usepackage{dcolumn}
\usepackage{bm}
\usepackage{amsfonts}
\usepackage{amsmath}
\usepackage{braket}
\usepackage[utf8]{inputenc}
\usepackage{hyperref}
\usepackage{color,soul}

\providecommand{\keywords}[1]
{
  \small	
  \textbf{\textit{Keywords---}} #1
}

\begin{document}

\title{Data-driven discovery of thermal illusions through latent-space geometry}

\author{Liyou Luo$^{1}$}
\author{Pengfei Zhao$^{1}$}
\author{Jensen Li$^{1,2}$}
\email{j.li13@exeter.ac.uk}

\affiliation{$^{1}$Department of Physics, The Hong Kong University of Science and Technology, Hong Kong, P. R. China.}
\affiliation{$^{2}$Department of Engineering, University of Exeter, Exeter, United Kingdom.}

\begin{abstract}
Illusion effects—where one object appears as another—arise from the non-uniqueness of physical systems, in which different material configurations yield identical external responses.
Conventional approaches, such as coordinate transformation, map equivalent configurations but provide only specific solutions, while analytical or numerical optimization methods extend these designs by minimizing scattering yet remain constrained by model assumptions and computational cost. Here, we exploit this non-uniqueness through a data-driven framework that uses a variational autoencoder to compress high-dimensional thermal-field data into a compact latent space capturing geometrical relations between configurations and observations. In this latent space, thermal illusion corresponds to finding configurations that minimize geometric distance to a target configuration, with thermal cloaking as a special case where the target is free space. Specifically, we demonstrate the concept in a cylindrical shell with anisotropic thermal conductivities enclosing a core of arbitrary conductivity, achieving robust thermal illusion and cloaking using only positive conductivities. Such a latent-space distance approach provides a refreshed perspective for achieving illusion and can be applied to inverse-design problems in other classical wave systems.
\end{abstract}

\keywords{invisibility cloak, illusion, machine learning, latent-space geometry, thermal conduction}

\maketitle

\section{Introduction}

Illusion~\cite{Pendry2006Controlling,Leonhardt2006Optical,Lai2009Illus,Liu2019Illusion,MachBatlle2020,Wu2025Reinforcement}, 
in which an object is made to appear as its background or as another object to outside observers, 
has attracted tremendous research interest over the past two decades. 
Originally demonstrated in optics, its concept has been extended to other domains—including acoustic waves~\cite{Chen2007Acoustic,Cummer2024Transformation}, 
fluid flow~\cite{Chen2024Multilayered}, and thermal conduction~\cite{Hu2018Illusion,Han2013Homogeneous}—to realize functional devices such as invisibility cloaks. 
The underlying principle is the equivalence of responses among distinct system configurations, allowing one configuration to camouflage another. 
A widely used method to design such equivalence is coordinate transformation, which constructs void spaces~\cite{Milton2006On,Xu2021Three} 
or complementary media~\cite{Lai2009Complementary} by exploiting the form-invariance of the governing equations~\cite{Rahm2008Design}. 
Although powerful, this construction can require extreme material parameters and is restricted to equivalence relations derivable from explicit transformations, 
thereby revealing only limited cases within the broader landscape of possible solutions.

In recent years, machine-learning techniques~\cite{LeCun2015Deep,Jordan2015Machine,Goodfellow2016Deep,Krizhevsky2012ImageNetNIPS} 
have provided a data-driven route to expand this design space. 
Supervised learning frameworks—such as inverse neural networks or tandem architectures—have been applied to metamaterial and illusion design~\cite{Qian2020Deep,Mirzakhanloo2020Active,Ji2022Design,Ahmed2021Deterministic,Fallah2023Developing}. 
These approaches typically attempt to learn an explicit inverse mapping from desired responses to material configurations, 
even though such mapping is intrinsically non-unique: multiple configurations can yield the same observable response. 
To ensure training convergence, they often restrict data or impose additional constraints~\cite{Kabir2008Neural,Ahmed2021Deterministic,Liu2018Training,So2019Simultaneous}, 
effectively avoiding this non-uniqueness rather than utilizing it. 
Here, we take the opposite approach. 
We show that this very non-uniqueness—previously regarded as an obstacle—is the key to achieving robust illusion. 
Taking thermal illusion as a specific example, by employing the unsupervised dimensional-reduction framework of the variational autoencoder~\cite{Kingma2014Auto,Doersch2016Tutorial}, 
we compress temperature-field data into a compact latent space that fully preserves the essential information 
while exposing the hidden geometric structure of equivalent responses. 
The reduced dimensionality reveals that the latent representation possesses fewer effective degrees of freedom than the number of physical parameters, 
signifying intrinsic non-uniqueness where different configurations produce identical responses. 
In this latent space, similarity between configurations is measured through geometric distance, 
allowing illusion and cloaking solutions to be identified efficiently. 
We demonstrate this concept by designing thermal illusion devices in the form of cylindrical shells with anisotropic conductivities 
that enclose a core of arbitrary conductivity.

\section{Results}

\begin{figure}[t]
  \centering
  \includegraphics[width=0.8\linewidth]{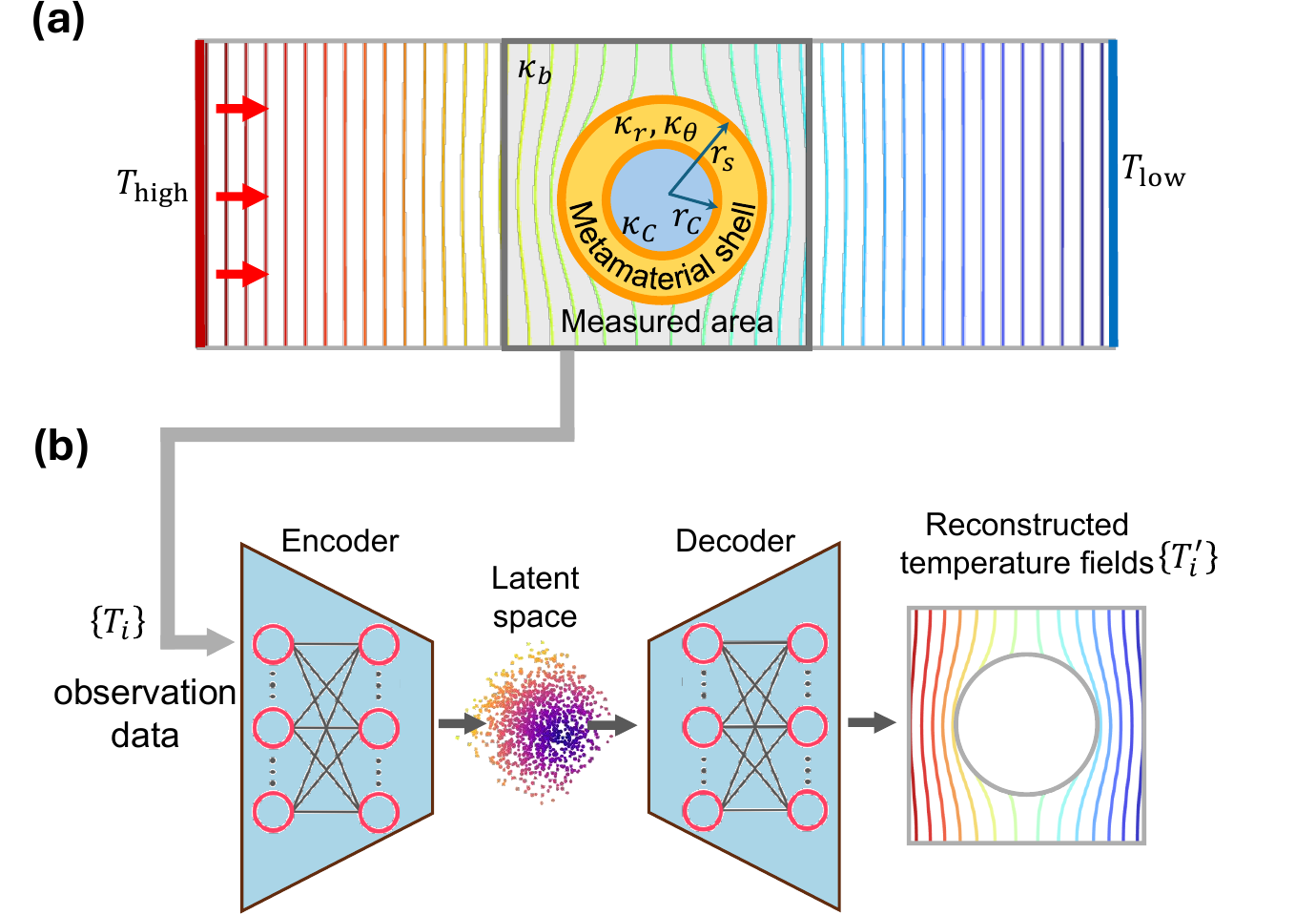}
  \caption{(a) Schematic of the thermal illusion framework. Heat flows from the left boundary at high temperature $T_{\mathrm{high}}$ to the right boundary at $T_{\mathrm{low}}$ through a cylindrical shell with anisotropic thermal conductivities $(\kappa_r, \kappa_\theta)$ enclosing a core of conductivity $\kappa_C$. The task is to find the material profile of the shell such that the overall system behaves like a core of target conductivity $\kappa_{\mathrm{target}}$ embedded in the background of conductivity $\kappa_b$, with thermal cloaking as the special case $\kappa_{\mathrm{target}}=\kappa_b$. Temperature fields are measured in the external region (gray area) outside the shell.  (b) Data-driven framework for analyzing equivalence in thermal responses. A variational autoencoder (VAE) encodes the measured temperature data into a compact latent space and decodes it to reconstruct the temperature fields. The latent representation provides a geometric measure of similarity between configurations, enabling the identification of cloaking and illusion solutions.}
  \label{fig:fig1}
\end{figure}

\medskip
\noindent\textbf{Data-driven thermal illusion framework.}
We demonstrate the proposed approach using a cylindrical shell designed for thermal illusion, as illustrated in Fig.~\ref{fig:fig1}(a). 
The system consists of a shell with anisotropic thermal conductivities $(\kappa_r, \kappa_\theta)$ surrounding a core of conductivity $\kappa_C$.
The anisotropic conductivities can, in principle, be realized using layers of isotropic materials in a metamaterial fashion \cite{Hu2018Illusion}, although this is not the focus of the present work for the implementation.
The outer and inner radii are set to $r_S = 3$ and $r_C = 1.5$ units, respectively, within a computational domain of total width $10$ units, where the upper and lower boundaries are thermally insulated. 
A high-temperature source $T_{\mathrm{high}}$ and a low-temperature sink $T_{\mathrm{low}}$ are applied at the left and right boundaries to establish a steady-state heat flow. 
The design goal is to determine the shell parameters $(\kappa_r, \kappa_\theta)$ so that the temperature field observed in the external region (gray area) matches that of a target configuration whose core has conductivity $\kappa_{\mathrm{target}}$ without the shell. 
The configuration should remain valid for any actual core conductivity $\kappa_C$, corresponding to a response robust against variations of the hidden object. 
Thermal cloaking appears as the special case $\kappa_{\mathrm{target}} = \kappa_b = 1$, where the system reproduces the homogeneous background response.

A direct search for such configurations based on the full temperature field would be computationally intensive due to the high dimensionality of the data. 
To overcome this, we employ a variational autoencoder (VAE) to compress the measured temperature fields into a compact latent space that preserves all relevant physical information while enabling efficient analysis. 
In this representation, each temperature field—originally containing over a thousand sampling points—is encoded as a single point in the latent space, where geometric distances directly reflect the similarity between thermal responses. 
Thus, the design task becomes finding a shell configuration $(\kappa_r, \kappa_\theta)$ whose latent representation lies closest to that of the target configuration defined by $\kappa_{\mathrm{target}}$.

\medskip
\noindent\textbf{VAE architecture and training.}
The VAE consists of an encoder that maps the temperature field $\mathbf{T}\in\mathbb{R}^{M}$ to a set of latent variables $\{z_i\}_{i=1}^{d}$, and a decoder that reconstructs $\mathbf{T}'\in\mathbb{R}^{M}$ from them, as showin in Fig. \ref{fig:fig1}(b). 
Each $z_i$ follows a Gaussian distribution with mean $\mu_i$ and variance $\sigma_i^2$, expressed by reparameterization.
\[
z_i = \mu_i + \epsilon \sigma_i, 
\qquad 
\epsilon \sim \mathcal{N}(0,1),\;\; i = 1,\dots,d,
\]
where $(\mu_1,\dots,\mu_d)$ and $(\sigma_1,\dots,\sigma_d)$ are outputs of the encoder. 
The network minimizes the $\beta$-VAE loss
\begin{equation}
\mathcal{L}_{\mathrm{VAE}}
=
\bigl\|\mathbf{T}-\mathbf{T}'\bigr\|_2^2
+
\beta\,
\sum_{i=1}^{d}
\left[
D_{\mathrm{KL}}\!\left(
\mathcal{N}\!\bigl(\mu_i,\sigma_i^{2}\bigr)
\,\big\|\, 
\mathcal{N}(0,1)
\right)
\right],
\label{eq:vae_loss}
\end{equation}
where the first term quantifies reconstruction accuracy and the second term (Kullback–Leibler divergence) regularizes the latent distribution toward a standard normal prior,
\[
D_{\mathrm{KL}}\!\left(
\mathcal{N}(\mu_i,\sigma_i^2) \,\big\|\, \mathcal{N}(0,1)
\right)
= \frac{1}{2}\left(\mu_i^{2}+\sigma_i^{2}-\ln\sigma_i^{2}-1\right).
\]
The hyperparameter $\beta$ controls the trade-off between compactness and reconstruction fidelity.

To generate the training data set, the material parameters are independently sampled as
$\kappa_r \sim U(0.1,10)$, $\kappa_\theta \sim U(0.1,10)$, and $\kappa_C \sim U(0.1,10)$ using uniform distributions. 
For each configuration, the steady-state temperature field is numerically simulated in \textsc{COMSOL Multiphysics}, recorded only in the measurement region excluding device region, and further flattened to a vector of length $M = 620$. Each simulation is performed with a background temperature gradient of -133.3 K/unit-length.
A total of $9{,}000$ temperature-field samples are generated from randomly selected configurations.
These samples are divided into training, validation, and testing sets in a 70\%:20\%:10\% ratio and trained using the Adam optimizer with a learning rate of $10^{-4}$ for 1000 epochs.
To identify the intrinsic dimensionality of the latent space, $\beta$ is gradually increased from $10^{-10}$ to $10^{-4}$: redundant latent dimensions are suppressed until a further increase later (a $\beta$ much larger than $10^{-4}$) causes a sharp increase in reconstruction error. Increasing $\beta$ encourages the use of fewer latent variables, allowing us to identify the minimal number required to faithfully represent the complex data—that is, the intrinsic dimensionality of the dataset.

\begin{figure}[t]
  \centering
  \includegraphics[width=0.7\linewidth]{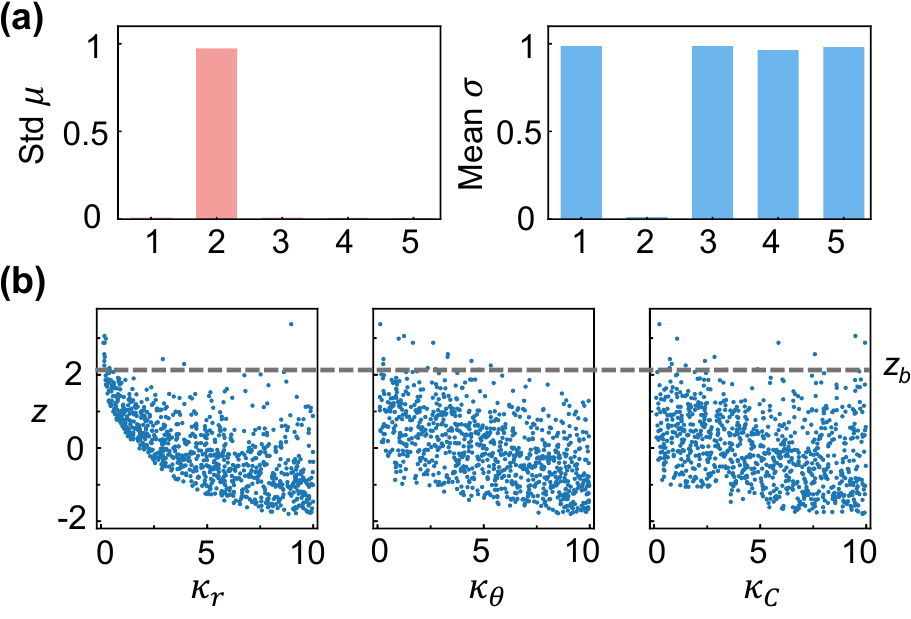}
  \caption{
(a) Statistical evaluation of latent variables. 
The standard deviation (std) of $\mu_i$ and the mean of $\sigma_i$ across the training dataset indicate whether a latent variable is meaningful. 
A meaningful variable corresponds to a well-defined generative factor, exhibiting low mean of $\sigma_i$ and a large std of $\mu_i$, while a meaningless one shows high $\sigma_i$ and small $\mu_i$ variations. 
In this case, only the second latent variable satisfies these criteria, suggesting that the temperature data are governed by a single effective degree of freedom. 
(b) Relationships between the meaningful latent variable $z$ and the material parameters $\kappa_r$, $\kappa_\theta$, and $\kappa_C$. 
Different parameter combinations yield the same $z$, revealing non-unique configurations that produce identical external temperature fields. 
The gray dashed line at $z = 2.14$ corresponds to the homogeneous background response, indicating potential cloaking configurations associated with low $\kappa_r$ values.
}
  \label{fig:fig2}
\end{figure}

\medskip
\noindent\textbf{Latent-space analysis.}
Each encoded temperature field corresponds to one point in the latent space, whose effective dimensionality is inferred from the statistical properties of the latent variables throughout the data set. 
We initialize the network with $d = 5$ latent variables and analyze the standard deviation of $\mu_i$ and the mean of $\sigma_i$, as shown in Fig.~\ref{fig:fig2}(a). 
A meaningful latent variable exhibits a low mean of $\sigma_i$ (indicating a well-defined value) and a large standard deviation of $\mu_i$ (indicating variation across samples)~\cite{Luo2025Inverse}. 
Only one variable meets these criteria, revealing that the temperature field can be described by a single dominant degree of freedom. 
We simply denote this variable as $z$ and plot its relation to the material parameters $(\kappa_r,\kappa_\theta,\kappa_C)$ in Fig.~\ref{fig:fig2}(b). 
The same $z$ can arise from multiple parameter combinations, confirming the existence of non-unique configurations that produce identical external temperature fields—precisely the condition for robust cloaking or illusion. 
In particular, $z=z_b=2.14$ (gray dashed line) corresponds to the homogeneous background (for the case of cloaking). We note that there is a large range of values in $\kappa_C$ spanning $0.1$--$10$, or a large range of values in $\kappa_\theta$ while only a limited range of low values in $\kappa_r$ in Fig. \ref{fig:fig2} can evaluate to the same $z_b$ (i.e. equivalent), where we can search for potential cloaking configurations.

\begin{figure}[t]
  \centering
  \includegraphics[width=0.7\linewidth]{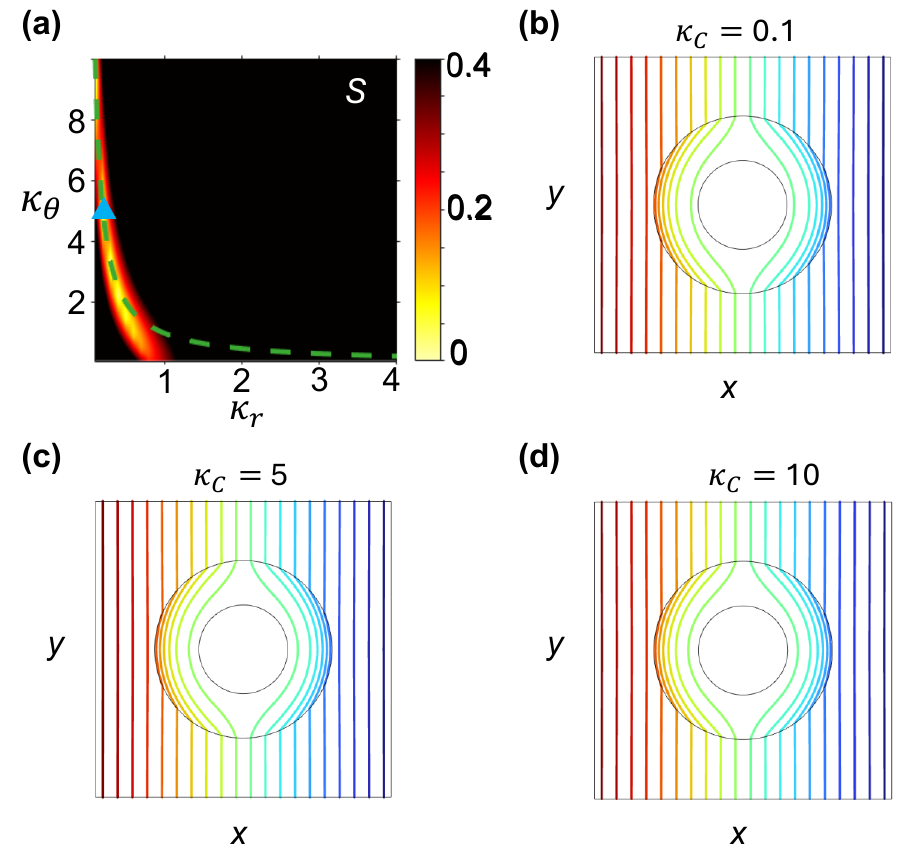}
  \caption{
(a) The cloaking metric $S$ as a function of $\kappa_r$ and $\kappa_\theta$, where smaller values (bright regions) indicate configurations that best reproduce the background response. Green dashed line indicates the condition $\kappa_r \kappa_\theta = 1$.
(b) Temperature contours of the selected configuration, marked by the blue triangle in (a) with $(\kappa_r, \kappa_\theta) = (0.2, 4.8)$, under three different core conductivities $\kappa_C = 0.1$, $5$, and $10$. 
The nearly vertical isothermal lines outside the shell demonstrate that the heat flux remains undisturbed, achieving the desired cloaking effect that makes the core appear as part of the background regardless of its thermal conductivity.
}
  \label{fig:fig3}
\end{figure}

\medskip
\noindent\textbf{Thermal cloaking.}
Having identified a single latent coordinate that captures the system’s thermal response (Fig.~\ref{fig:fig2}), we now use this compact representation to design robust cloaking configurations.  
Up to this point, we have established the mapping $z = Z(\mathbf{T})$ from the encoder, which converts a temperature field into its latent variable.  
Next, we construct a function that expresses the latent variable $z$ directly in terms of the physical parameters,
\[
z = Z(\kappa_r, \kappa_\theta, \kappa_C),
\]
thereby linking each material configuration to its latent representation derived from the corresponding external temperature field.  
To achieve this, we train a small supervised regression network using the same dataset, with conductivities as inputs and latent variables as targets, enabling interpolation of latent representations beyond the original training samples.  
It is important to note that the computational complexity is primarily concentrated in the VAE, while the regression network—operating within the reduced latent space—is comparatively lightweight and efficient.

Because $z$ encodes the observable temperature profile, the Euclidean distance between two latent points directly quantifies the difference between their external thermal responses. 
The search for cloaking configurations can therefore be reformulated as a geometric problem in latent space. 
We define a cloaking metric $S$ to evaluate how closely a given shell configuration $(\kappa_r, \kappa_\theta)$ reproduces the background response when averaged over variations in the core conductivity $\kappa_C$:
\begin{equation}
S(\kappa_r, \kappa_\theta) =
\mathrm{mean}_{\{\kappa_C\}}
\!\left(
\bigl|\, Z(\kappa_r, \kappa_\theta, \kappa_C) - z_b \,\bigr|
\right),
\label{eq:S_metric}
\end{equation}
where $z_b = Z(\kappa_b, \kappa_b, \kappa_b)$ is the latent coordinate of the homogeneous background. 
A smaller $S$ indicates that the shell produces an external temperature field nearly indistinguishable from the background, independent of changes in $\kappa_C$.

\begin{figure}[t]
  \centering
  \includegraphics[width=0.7\linewidth]{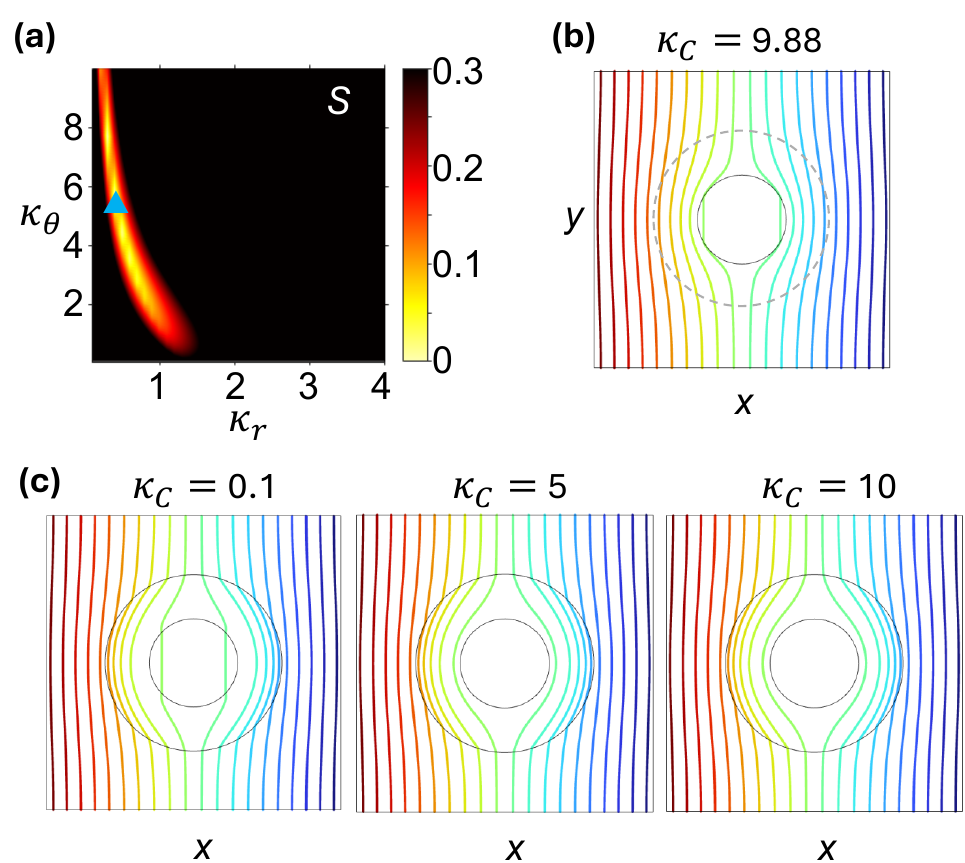}
  \caption{(a) The illusion metric $S(\kappa_r,\kappa_\theta;\kappa_{\mathrm{target}}{=}9.88)$ shown as a colormap, where bright regions represent configurations that closely reproduce the response of a bare core with $\kappa_{\mathrm{target}}=9.88$ embedded in the background. (b) The temperature field of the bare core with $\kappa_{\mathrm{target}}=9.88$ in the background. (c) The selected configuration (marked by the blue triangle in (a)) with $(\kappa_r,\kappa_\theta)=(0.397,5.644)$ produces nearly identical temperature patterns to (b), even when the actual core conductivity varies across $\kappa_C=0.1$, $5$, and $10$, misleading external observers to perceive a bare core of $\kappa_{\mathrm{target}}=9.88$.}
  \label{fig:fig4}
\end{figure}

The computed map of $S(\kappa_r,\kappa_\theta)$ is shown in Fig.~\ref{fig:fig3}(a). 
The bright ridge corresponds to the region of smallest $S$, revealing potential cloaking configurations. 
For homogeneous anisotropic shells inspired by transformation thermodynamics, a common simplification is to impose a constant determinant of the conductivity tensor, i.e., $\kappa_r\kappa_\theta=\kappa_b^2$ (with $\kappa_b=1$ here, indcated by green dashed line in \ref{fig:fig3}(a)), as used in Ref.~\cite{Guenneau2012Transformation,Han2013Homogeneous}. 
Along our optimal ridge in the $(\kappa_r,\kappa_\theta)$ map, we find that this relation is well approximated in the strongly anisotropic regime (very small $\kappa_r$). 
However, as $\kappa_r$ increases (reduced anisotropy), the optimal solutions deviates from the constant–product condition, as discovered by the current approach. 
A representative case, $(\kappa_r,\kappa_\theta)=(0.2,4.8)$, marked by the blue triangle, is verified by direct simulation. 
The temperature contours for $\kappa_C=0.1$, $5$, and $10$, shown in Fig.~\ref{fig:fig3}(b),(c) and (d) respectively, exhibit nearly identical patterns: the isothermal lines outside the shell remain vertical and parallel, indicating that the heat flux flows undisturbed as in the background. 
Meanwhile, the temperature within the core is nearly uniform, corresponding to an almost zero temperature gradient and hence negligible thermal current. 
We also estimate an effective temperature gradient using only the outer circumference of the device ($r = r_s$), treating the enclosed region as a black box. 
The gradient is calculated from 
\(\int x T \, d\theta \, / \, (\pi r_S^2)\), 
where $\theta$ is the angular coordinate. 
The resulting values, $-135.0$, $-134.7$, and $-134.7$~K per unit length for the three cases in Fig.~\ref{fig:fig4}(b),(c) and (d), 
closely match the target value of $-133.3$~K per unit length as if it is just a background medium. We can also evaluate such an effective temperature gradient at $r=r_C$ (by changing $r_S$ to $r_C$ in the definition), we obtain $-16.4$, $-3.0$ and $-1.6$ K/unit-length the the three cases, i.e. nearly zero temperature gradient. 
These results demonstrate that the cloak effectively insulates the core from external heat flow while maintaining a background-like response to outside observers, achieving both concealment and protection for temperature-sensitive components.

\medskip
\noindent\textbf{Thermal illusion.}
An advantage of the current latent-space distance approach is its natural extensibility to realizing illusion effects beyond cloaking, where the system mimics the thermal response of an arbitrary target object rather than that of the background.
This generalization can be achieved simply by replacing the reference latent coordinate $z_b$ in Eq.~(\ref{eq:S_metric}) with that of a target configuration, $z_{\mathrm{target}} = Z(\kappa_b, \kappa_b, \kappa_{\mathrm{target}})$. 
The same metric is now adapted to
\begin{equation}
S(\kappa_r,\kappa_\theta;\kappa_{\mathrm{target}}) = 
\mathrm{mean}_{\{\kappa_C\}}
\bigl(
|\,z(\kappa_r,\kappa_\theta,\kappa_C) - z_{\mathrm{target}}\,|
\bigr).
\label{eq:illusionMetric}
\end{equation}
Such illusion metric measures how closely a shell configuration $(\kappa_r,\kappa_\theta)$ reproduces the target response while remaining robust to variations in the actual core conductivity $\kappa_C$. 
Thermal cloaking corresponds to the special case where $\kappa_{\mathrm{target}}=\kappa_b$.

As an example, we consider $\kappa_{\mathrm{target}} = 9.88$ and design a shell that makes any embedded core appear as a homogeneous inclusion with this conductivity.  
Without retraining the VAE, we use the same latent-variable function $z = Z(\kappa_r, \kappa_\theta, \kappa_C)$ to evaluate the illusion metric in Eq.~\ref{eq:illusionMetric}.
Figure~\ref{fig:fig4}(a) plots $S(\kappa_r,\kappa_\theta;\kappa_{\mathrm{target}}{=}9.88)$, revealing a bright region that indicates potential illusion configurations. 
We select one case, $(\kappa_r,\kappa_\theta)=(0.397,5.644)$, marked by the blue triangle. 
The reference temperature field of the bare core with $\kappa_{\mathrm{target}}=9.88$ is shown in Fig.~\ref{fig:fig4}(b), and the corresponding results for the selected shell under $\kappa_C=0.1$, $5$, and $10$ are presented in Fig.~\ref{fig:fig4}(c). 
The nearly identical temperature contours in \ref{fig:fig4}(b) and (c) confirm that the shell successfully misleads external observations, producing the same apparent response as the target object despite variations in the hidden core. We estimate an effective temperature gradient using only the outer circumference of the device ($r = r_s$) defined previously, treating the enclosed region as a black box. The resulting values of temperature gradient are $-105.2$, $-104.2$, and $-104.0$~K per unit length for the three cases in Fig.~\ref{fig:fig4}(c), 
closely match the target value of $-103.9$~K per unit length in Fig.~\ref{fig:fig4}(b), confirming the effectiveness of the illusion.
The current data-driven design framework, grounded in the geometric structure of latent space, not only provides thermal illusion by using positive conductivities but also unifies cloaking and illusion within a single framework and offers a practical route toward experimentally realizable thermal metamaterials.

\section{Conclusion}

We have introduced a data-driven framework for realizing thermal cloaking and illusion effects by exploiting the intrinsic non-uniqueness of the mapping between physical configurations and their observable responses. 
Instead of eliminating this non-uniqueness, our approach leverages it as a design resource, using a variational autoencoder to compress high-dimensional temperature-field data into a compact latent representation where geometrical distances directly quantify similarity of responses. 
Within this space, configurations that yield equivalent or target-mimicking behavior emerge naturally as clusters or nearby points, providing an efficient and interpretable route for identifying robust solutions.
Demonstrated through a cylindrical shell with anisotropic thermal conductivities, the framework unifies cloaking and illusion under the same latent-space metric: cloaking corresponds to matching the background response, while illusion generalizes this to any chosen target. 
By revealing that multiple material configurations can produce indistinguishable observations, the method offers a geometric perspective for inverse design that extends beyond thermal conduction to other wave and transport phenomena, enabling scalable and physically interpretable discovery of metamaterial functionalities.






\bibliography{bib}

\end{document}